\documentclass{article}

\usepackage{longtable}
\usepackage{subcaption}
\usepackage{graphicx}
\usepackage[utf8]{inputenc} % allow utf-8 input
\usepackage[T1]{fontenc}    % use 8-bit T1 fonts
\usepackage{hyperref}       % hyperlinks
\usepackage{url}            % simple URL typesetting
\usepackage{booktabs}       % professional-quality tables
\usepackage{amsfonts}       % blackboard math symbols
\usepackage{nicefrac}       % compact symbols for 1/2, etc.
\usepackage{microtype}      % microtypography
\usepackage{xcolor}         % colors
\usepackage{biblatex}

\DeclareUnicodeCharacter{22C6}{ }
\DeclareUnicodeCharacter{2264}{ }
\DeclareUnicodeCharacter{223C}{ }
\usepackage[final, nonatbib]{neurips_2021_ml4ps}
\title{Photometric Redshifts for Cosmology: Improving Accuracy and Uncertainty Estimates Using Bayesian Neural Networks}

\author{
  Evan Jones\\
University of California, Los Angeles\\
  \texttt{evan.jones@astro.ucla.edu} \\
   \And
  Tuan Do \\
  University of California, Los Angeles\\
  \texttt{tdo@astro.ucla.edu} \\
   \And
  Bernie Boscoe \\
  Occidental College\\
  \texttt{boscoe@oxy.edu} \\
  \And
  Yujie Wan\\
  University of California, Los Angeles\\
  \texttt{wanyujie0606@gmail.com} \\
  \And
   Zooey Nguyen\\
  University of California, Los Angeles\\
  \texttt{zooeyn@ucla.edu} \\
  \And
  Jack Singal\\
  University of Richmond\\
  \texttt{jsingal@richmond.edu} \\
}

\begin{document}

\maketitle

\begin{abstract}
We present results exploring the role that probabilistic deep learning models can play in cosmology from large scale astronomical surveys through estimating the distances to galaxies (redshifts) from photometry. Due to the massive scale of data coming from these new and upcoming sky surveys, machine learning techniques using galaxy photometry are increasingly adopted to predict galactic redshifts which are important for inferring cosmological parameters such as the nature of dark energy. Associated uncertainty estimates are also critical measurements, however, common machine learning methods typically provide only point estimates and lack uncertainty information as outputs. We turn to Bayesian neural networks (BNNs) as a promising way to provide accurate predictions of redshift values. We have compiled a new galaxy training dataset from the Hyper Suprime-Cam Survey, designed to mimic large surveys, but over a smaller portion of the sky. We evaluate the performance and accuracy of photometric redshift (photo-z) predictions from photometry using machine learning, astronomical and probabilistic metrics. We find that while the Bayesian neural network did not perform as well as non-Bayesian neural networks if evaluated solely by point estimate photo-z values, BNNs can provide uncertainty estimates that are necessary for cosmology.

\end{abstract}

\section{Introduction}
As large astronomical surveys come online in the next few years both from the ground with LSST and the Vera Rubin Telescope and in space with Euclid, many researchers are turning to machine learning to handle the exponentially increasing influx of data. However, common machine learning methods often only provide point estimates and do not generally provide accurate confidence intervals for specific predictions \cite{graff_skynet_2014, jones_analysis_2017,carrasco_photometric_2015}. Accurate uncertainty estimates, whether from machine learning or elsewhere, are critical for measurements from these surveys because these measurements and uncertainties are not the end goals; rather, they are subsequent inputs into inference to constrain models of our Universe. 

One crucial goal of these surveys is to determine expansion history of the universe and with it the parameters describing dark energy.  This determination relies ultimately on accurately and precisely measuring the redshifts of hundreds of millions of galaxies.  Spectroscopically measuring galaxy redshifts, where the light is split into hundreds of small bins of wavelength, is time consuming and practically impossible for the necessary sample size to constrain cosmological parameters.  Instead of measuring detailed galaxy spectra to determine redshift, one can take images of galaxies in a few large bins of wavelength (photometry). While galaxy photometry contains information about redshift, the observed variation between the intrinsic properties of galaxies makes it difficult to model a-priori. Astronomers have adopted data driven approaches using machine learning methods for redshift estimates using photometry \cite{schuldt_photometric_2021, aihara_second_2019, carrasco_kind_tpz_2013, tanaka_photometric_2015, jones_tests_2020, wyatt_outlier_2021}.

In this work, we investigate photometric redshift (photo-z) estimation using Bayesian neural networks (BNN), a type of probabilistic neural network (NN) \cite{jospin_hands-bayesian_2020}. Probabilistic neural networks, conceptualized in the 1990s \cite{specht_probabilistic_1990}, have previously been limited in their ability to process the size of data required for performing photo-z estimation for large-scale surveys, because of the complexity of their computation. However, recent breakthroughs in conceptual understanding and computational capabilities (e.g. \cite{filos_systematic_2019, dusenberry_analyzing_2020}) now make probabilistic deep learning possible for cosmology. Probabilistic deep learning such as BNN has many advantages compared to traditional neural networks, including better uncertainty representations, better point predictions, and offers better interpretability of neural networks because they can be viewed through the lens of probability theory. In this way we can draw upon decades of development in Bayesian inference analyses. We compare the performance of a BNN to a fully-connected non-probabilistic NN and evaluate the accuracy of the confidence intervals for the probabilistic predictions.  To our knowledge this is the first application of BNN for photo-z estimation. We compare the NN to BNN in order to assess the effect of incorporating Bayesian statistics to photo-z estimation; we intentionally compose both models to be as similar as possible.

\section{Data and Methods}

\subsection{Data: Galaxy observations}
%Describe the data and how we made the catalog:
\begin{figure}[!h]
\begin{subfigure}{1\textwidth}
\resizebox{\hsize}{!}
{\includegraphics{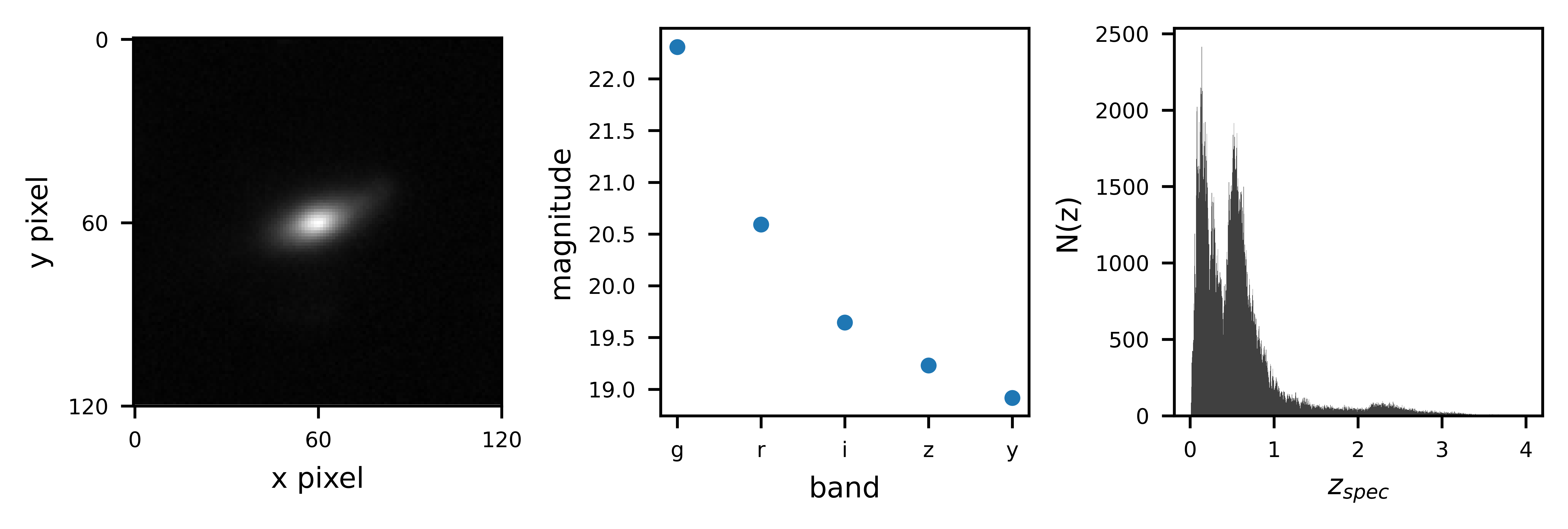}}
\label{z_rems}
\end{subfigure}
\caption{Left: typical galaxy (z = 0.48) image in the $i$-band. Middle: five-band photometry for the same galaxy. Right: N(z) distribution for the dataset discussed in \S \ref{data} For the photo-z determinations in this work we use training and testing sets consisting of 229,120 and 28,640 galaxies respectively.}
\label{fig1}
\end{figure}
For the analysis in this work we compile a dataset intended to approximate the data produced by future large-scale deep surveys for photo-z estimation \cite{the_lsst_dark_energy_science_collaboration_lsst_2021}.  We use the Hyper-Suprime Cam (HSC) Public Data Release 2 (PDR2) \cite{aihara_second_2019}, which is designed to reach similar depths as LSST but over a smaller portion of the sky. We crossmatched galaxy photometry from HSC with the HSC collection of publicly available spectroscopic redshifts  \cite{lilly_zcosmos_2009},\cite{bradshaw_high-velocity_2013, mclure_sizes_2013}, \cite{skelton_3d-hst_2014, momcheva_3d-hst_2016},\cite{le_fevre_vimos_2013}, \cite{garilli_vimos_2014}, \cite{liske_galaxy_2015}, \cite{davis_science_2003, newman_deep2_2013}, \cite{coil_prism_2011, cool_prism_2013} using the galaxies' sky positions (d < 1 arcsecond). We use data quality cuts similar to \cite{nishizawa_photometric_2020} and \cite{schuldt_photometric_2021} (see \footnote{\url{https://doi.org/10.5281/zenodo.5528827}} for full list). We use the spectroscopic redshift values as the ground truth for training and evaluation. We also select only one set of g,r,i,z,y measurements per galaxy. In total, our data consists of  286,401 galaxies with broad-band g,r,i,z,y photometry from the HSC PDR2 survey and spectroscopic redshifts.  The majority of galaxies in our sample lies between redshift of 0.01 and 2.5 (see N(z) in Fig. 1). We use 80\% for training, 10\% for validation, and 10\% for testing. 

\label{data}

\subsection{Network architectures}
We compare the performance of a NN and BNN using a similar architecture. Both the NN and the BNN are implemented in TensorFlow and have five input nodes for photometry with four hidden layers (200 nodes per layer with rectified linear activation function). The networks also have a skip connection between the input nodes and the final layer. The NN has an output node to produce a single point estimate photo-z prediction. The BNN has a final output node that produces a mean and standard deviation assuming a Gaussian distribution for each photo-z prediction. For the BNN we use a negative log likelihood loss function with RMS error as the metric while the NN uses a mean absolute error loss function. Both models use the Adam optimizer.  We train using an AMD Ryzen Threadripper PRO 3955WX with 16-Cores and NVIDIA RTX A6000. Training and evaluation runtimes are typically under 30 minutes.

While optimal hyperparameters for each network such as the number of nodes per layer and number of layers differed slightly between the BNN and NN, we find the difference in performance is negligible and use similar architectures when possible for the sake of comparison.  We choose the negative log-likelihood loss function for the BNN because it has been shown to be more effective than MAE for probabilistic NNs \cite{lakshminarayanan_simple_2017}.
\begin{figure}[!h]
\begin{subfigure}{1\textwidth}
\resizebox{\hsize}{!}
{\includegraphics{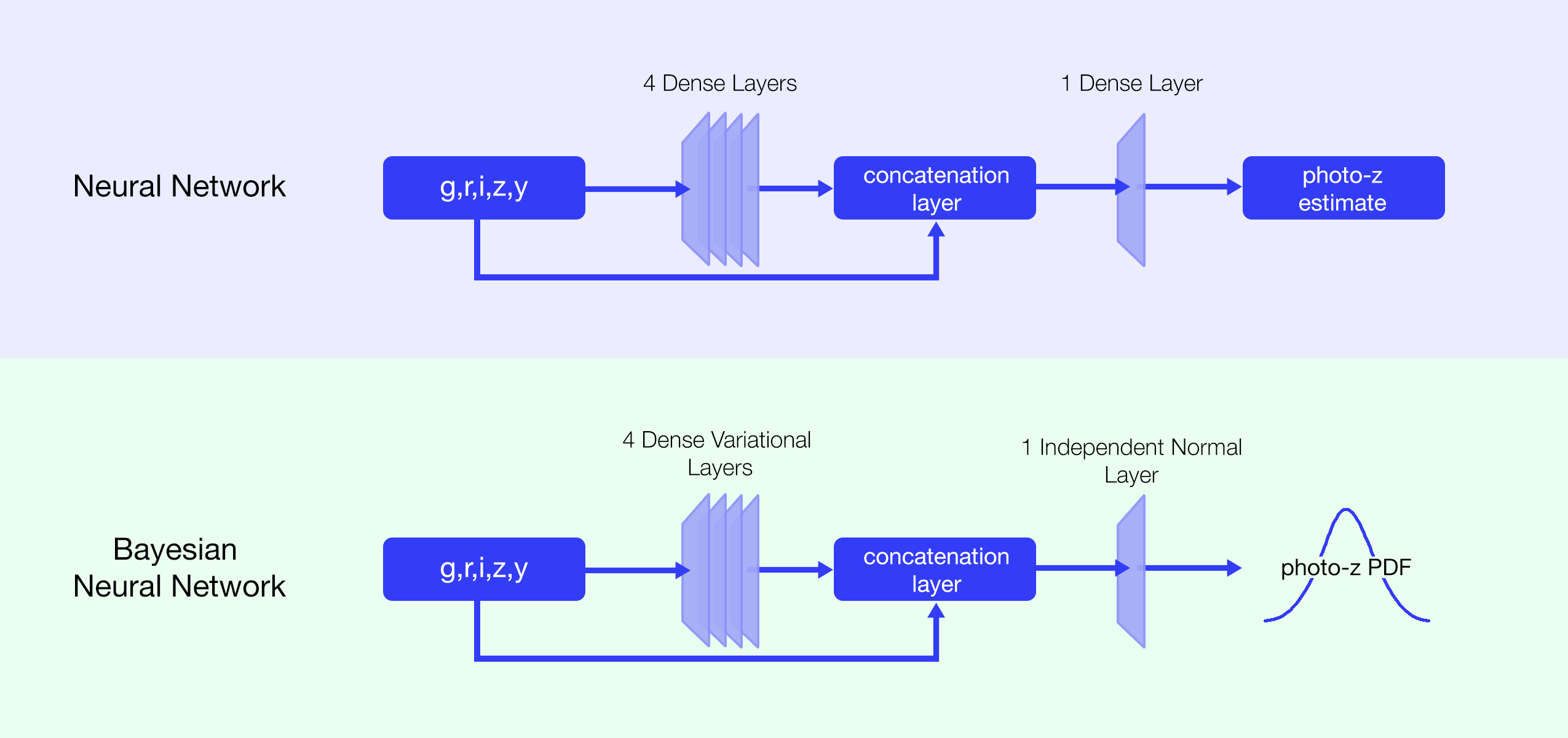}}
\label{z_rems}
\end{subfigure}
\caption{Top: NN architecture. Bottom: BNN architecture. The inputs for both networks are five-band photometry in the g,r,i,z,y filters. The output for the NN is a discrete photo-z estimate while the output for the BNN is a photo-z PDF, which we sample to obtain a photo-z estimate. We assume Gaussianity in the creation of the photo-z PDF, so a photo-z uncertainty is produced by the standard deviation of the PDF.}
\label{fig1}
\end{figure}

\label{network_architectures}

\begin{figure}[!h]
\begin{subfigure}{0.99\textwidth}
\resizebox{\hsize}{!}
{\includegraphics{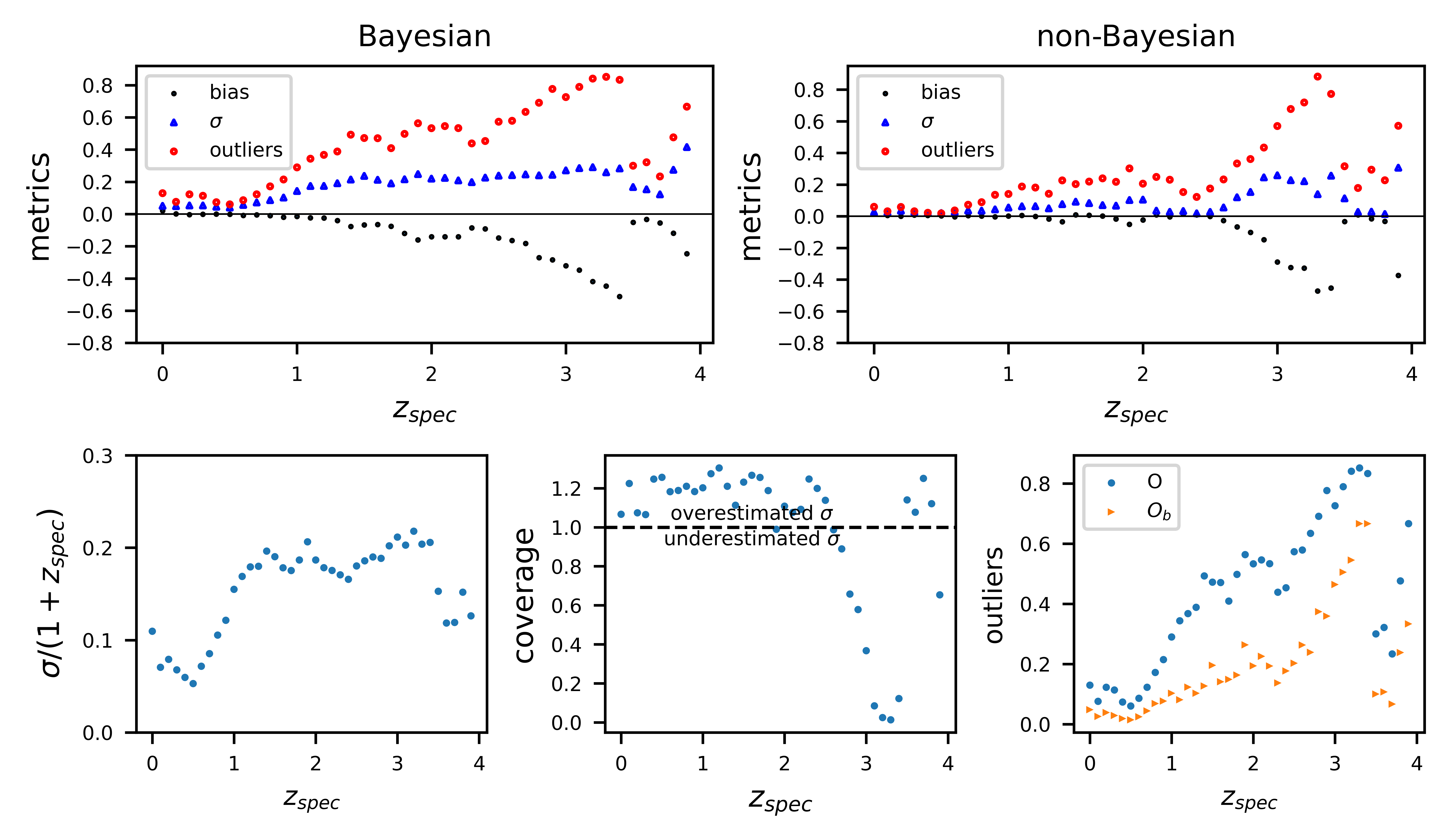}}
\label{z_rems}
\end{subfigure}
\caption{Results here are binned into groups of 0.1 redshift and averaged over each bin. Top row - Model performance with the NN and BNN versus redshift.  Bottom row - Left: photo-z uncertainty produced by the BNN scaled by true redshift. Middle: coverage of photo-z predictions with the BNN as discussed in \S \ref{metrics}. Right: outliers produced by the NN and BNN as defined in Eqs. 1 and 2.}
\label{fig2}
\end{figure}

\label{architectures}
\subsection{Metrics}
%The dark energy survey requires (insert) outliers and ( )rms . 
To measure model performance we evaluate predictions using the metrics in Table 2. We define ``outliers'' in Eq. 1, where $z_{phot}$ and $z_{spec}$ are the estimated photo-z and actual (spectroscopically determined) redshift of the galaxy. An advantage of the BNN is that the model naturally outputs an uncertainty for each photo-z prediction; using the associated uncertainties we can consider an additional quality metric defined in Eq. 2, where the uncertainty $\sigma$ is the standard deviation of the photo-z PDF produced by the BNN. The RMS photo-z error in a determination is given by a standard definition in Eq. 3, where $n_{gals}$ is the number of galaxies in the evaluation testing set and $\Sigma_{gals}$ represents a sum over those galaxies. Bias and dispersion are defined in Eqs. 5 and 6, where MAD is the median absolute deviation. We follow \cite{tanaka_photometric_2018} and define a loss function in Eq. 7 to characterize the point estimate photo-z accuracy with a single number, where we use $\gamma = 0.15$.

Finally, a key metric in assessing the performance of the BNN is `coverage', which we use as a metric for determining whether we have accurate uncertainties. Coverage is the fraction of galaxies that have a spectro-z within their 68$\%$ confidence interval. Ideally, 68$\%$ of evaluated galaxies should have true spectro-zs within their 68$\%$ confidence interval. If more than 68$\%$ of evaluated galaxies have spectro-zs within their 68$\%$ confidence interval, the galaxies are considered `over-covered' because their photo-z uncertainties are too large. The same logic applies for `under-covered' galaxies. 

\label{metrics}
\begin{table}
\caption{Metrics used to assess model performance.}
\centering
\begin{tabular}{lrrr}
\hline
\hline
Point Metrics & & Probabilistic Metrics\\
\hline
\hline
Outlier & $O: {{\vert z_{phot}-z_{spec} \vert} \over {1+z_{spec}}} > .15$ (1) & Bayesian Outlier $O_b: {{\vert z_{phot}-z_{spec} \vert - \sigma} \over {1+z_{spec}}} > .15$ (2) \\
RMS error & $  \sqrt { {{1} \over {n_{gals}}} \Sigma_{gals} \left( {{ z_{phot}-z_{spec} } \over {1+z_{spec}}} \right) ^2 } $ (3) & coverage: $ \displaystyle\sum_i^{n_{gals}} \frac{(\bar{z}_{pdf,i} - z_{spec,i}) < z_{\sigma,i}}{n_{gals}} $ (4)\\
bias & $b = {{ z_{phot}-z_{spec} } \over {1+z_{spec}}}$ (5)  \\
MAD &   Median($|\Delta z - $Median$(\Delta z_i)|)$ (6)\\
loss & $  L(\Delta z) = 1 - \frac{1}{1+(\frac{\Delta z}{\gamma})^2} $ (7)\\

\hline
\end{tabular}
\end{table}

\section{Results}

We compare the performance of the NN and BNN on the data discussed in \S \ref{data} in Table 2. The percentage of point-source outlier predictions as defined by Eq. 1 differ for the NN and BB. For the NN, only 6.5\% of points lie outside of the lines as outliers for the NN and 17.3\% for the BNN. With the modified outlier metric (Eq. 2) discussed in \S \ref{metrics} we obtain $O_b = 6.3\%$. Fig. 2 contains results from an example determination with a BNN and non-Bayesian NN, where results are divided into bins of size $z = 0.1$ and averaged. We note that both models generally perform worse at higher redshifts, which is due in large part to the reduced signal to noise for distant dim sources and also the disproportionate number of high redshift sources (z > 2.5) compared to low redshift sources. The latter point is an unavoidable attribute of similar datasets. We note that low redshifts are generally over-covered, indicating that their photo-z uncertainties are over-estimated, while the photo-z uncertainties of high redshift galaxies are generally under-estimated.

While our goal in this work is to compare the two types of NNs, we provide a comparison to LSST requirements \cite{the_lsst_dark_energy_science_collaboration_lsst_2021} (Table 2) for reference. The NN meets the LSST goal for outlier rate and bias. We believe both models could be further optimized for these requirements.

\begin{table} 
\caption{Comparison of BNN to NN performance averaged over all evaluation galaxies for a sample determination. We include LSST science requirements for reference when possible.}
\centering
\begin{tabular}{lrrrrrrrr}
\hline
\hline
Network & $O$ & $O_b$ & RMS & $|b|$ & MAD & $L(\Delta z)$ & coverage & ${\sigma}/(1+z_{spec})$ \\
\hline
BNN & 0.173 & 0.063 & 0.225 & 0.007 & 0.074 & 0.22 & 0.78 & 0.005 \\
NN  & 0.065 & - &0.174 & 0.002 & 0.023 & 0.095 & - & -\\
\hline
LSST Req. & < 0.15 & - & - & < 0.003 & < 0.02 \\
\end{tabular}
\end{table}
\section{Discussion}

Compared to the non-probabilistic NN, the BNN has the advantage of producing uncertainty constraints on every prediction, which are necessary for using photo-z estimation as a probe of cosmological parameters. Based on the results in this work, the BNN has the disadvantage of generally producing worse point estimates compared to the NN, however, pairing photo-z predictions with uncertainties provides a more robust look into the quality of photo-z predictions. We note that the BNN in this work was designed with the intention to closely resemble the non-probabilistic NN, and therefore our findings may not generalize to other BNN models. Optimal photo-z performance from the BNN may require significant model adjustments, which is an on-going study. Fig. 2 (right) visualizes the $O$ and $O_b$ rates per redshift bin; as expected, the number of outliers decreased when considering the photo-z uncertainty. The uncertainties produced by the BNN are larger than expected for $0 < z < 2.5$ and are underestimated in $ 2.5 < z < 4$. This is not necessarily a flaw inherent to a BNN, but it is worth investigating how to develop better uncertainty estimates with broad-band photometry. It is possible that one source of the over-estimation of photo-z uncertainties results from a disparity between the complexity present in the band magnitudes compared to the BNN model; we use five photometric band fluxes paired with a single spectroscopic redshift per galaxy for training, while the model parameters optimized during the training process of a NN can easily reach into the thousands. In a future work we will apply galaxy photometric images to a Bayesian convolutional neural network, which we believe will enhance the information present in the band fluxes.

\printbibliography
\end{document}